\documentclass[twocolumn,prb,amsmath,amssymb,citeautoscript,floatfix]{revtex4}

\usepackage{graphicx,color}

\usepackage{amsmath,amssymb}

\begin{document}

\title{Electron transport in high-entropy alloys: \\
Al$_{x}$CrFeCoNi as a case study.
}

\author{J. Kudrnovsk\'y}\email{kudrnov@fzu.cz} \affiliation{Institute of Physics ASCR, Na Slovance 2, CZ-182 21 Praha 8,
Czech Republic}

\author{V. Drchal} \affiliation{Institute of Physics ASCR, Na Slovance 2, CZ-182 21 Praha 8, Czech
Republic}

\author{F. M\'aca} \affiliation{Institute of Physics ASCR, Na Slovance 2,
CZ-182 21 Praha 8, Czech Republic}

\author{I. Turek}
\affiliation{Institute of Physics of Materials ASCR, \v{Z}i\v{z}kova 22, CZ-616 62 Brno, Czech Republic}

\author{S. Khmelevskyi}
\affiliation{Center for Computational Materials Science,
Institute for Applied Physics, Vienna University of Technology,
Wiedner Hauptstrasse 8, A-1040 Vienna, Austria}
\date{\today}

\begin{abstract}
The high-entropy alloys Al$_{x}$CrFeCoNi exist over a broad
range of Al concentrations ($0 < x < 2$).
With increasing Al content their structure is changed from the fcc
to bcc phase.
We investigate the effect of such structural
changes on transport properties including the residual
resistivity and the anomalous Hall resistivity.
We have performed a detailed comparison of the first-principles
simulations with available experimental data.
We show that the calculated residual resistivities for all studied
alloy compositions are in a fair agreement with available
experimental data as concerns both the resistivity values and
concentration trends.
We emphasize that a good agreement with experiment was obtained
also for the anomalous Hall resistivity.
We have completed study by estimation of the anisotropic
magnetoresistance, spin-disorder resistivity, and Gilbert damping.
The obtained results prove that the main scattering mechanism is
due to the intrinsic chemical disorder whereas the effect of spin
polarization on the residual resistivity is appreciably weaker.
\end{abstract}

\maketitle

\section{Introduction}

\label{Intr}

The high-entropy alloys (HEA), the multicomponent crystalline
alloys, often also called multi-principal element alloys
have attracted a quite significant and  growing interest in
the last decade.
Of the vast existing literature we just mention a recent book
\cite{hea-book}, a critical review \cite{hea-rev},
and an overview of possible theoretical approaches \cite{mupei}.
The high entropy of mixing of these multicomponent alloys suppresses
the formation of ordered intermetallic compounds leading to well
disordered phases with simple lattice structures such as the
face-centered cubic (fcc) or the body-centered cubic (bcc) ones.
Magnetic HEA's that consist of magnetic 3$d$-elements are particularly
interesting.
A typical example is the so-called quinary Cantor alloy (CrMnFeCoNi)
consisting of equiconcentration disordered mixture of magnetic
Cr, Mn, Fe, Co, and Ni elements with an fcc structure.
Such alloy offers a richness of magnetic properties depending on
the sample preparation \cite{cantor}.
The structural change from the fcc to bcc phase was predicted by
{\it ab initio} molecular dynamics simulations \cite{cantor2}
for Cantor-like alloys with Cr substituted by another metallic
element with formula CoFeMnNiX (X = Al, Ga, and Sn).

By doping with $sp$-elements, which influences carrier
concentration in the conduction band and thus both the magnetic and
transport properties, and even the alloy structure one can search
for new functional properties.
A typical example of such alloy is Al$_{x}$CrFeCoNi alloy
\cite{axcfcn,axcfcn2} with $x$ ranging from $x$=0 to $x$=2.
In particular, alloying with increasing Al content stabilizes
the bcc phase from the original fcc phase of quaternary CrFeCoNi
alloy.
Another interesting property, also present in the Cantor alloy,
is a large residual resistivity of the order of 100~$\mu\Omega$cm
which is in a striking contrast to a much smaller resistivity of
the binary fcc NiFe or fcc NiCo counterparts.
The large values of the residual resistivity in HEA's are caused
by strong scattering on the intrinsic chemical disorder.
In the present work we apply the alloy-specific
first-principles methodology based on the Kubo-Greenwood formula
\cite{sigma} which was successfully used for binary alloys
\cite{rsigma,vertex} also to HEA's in order to estimate the intrinsic
contribution to the resistivity and compare them to available
experimental data.

Contrary to the experimental and theoretical studies of structural
and thermodynamical properties of HEA's the studies of electronic transport are
very rare \cite{hea-rev}. The transport properties, together with the electronic
structure are among the most important material properties.
The theoretical tools for the resistivity study are more complicated and
not so broadly available as electronic structure codes focused on total
energies, electron densities and magnetic moments. Recently, theoretical
transport studies of the Cantor fcc CrMnFeCoNi alloy \cite{hea-ebert} and
of a related medium-entropy fcc NiCoMn alloy \cite{Mu-PRMater} appeared
which studied various possible scattering mechanisms contributing to
the residual resistivity.
However, theoretical investigation of the effect of Al-doping on
the resistivity, as well as of the role of different structures (bcc, fcc)
for electron transport in AlCrFeCoNi, is still missing.
Moreover, in addition to residual resistivities also the anomalous
Hall resistivity (AHR) for both structure phases was determined experimentally
\cite{axcfcn,axcfcn2}.
Therefore, these alloys are an obvious choice for ab initio based studies of
transport properties in HEA's containing $sp$-elements.
In the present study, also the anisotropic magnetoresistance (AMR),
the spin-disorder resistivity and the Gilbert damping for both structures
and typical Al concentrations are calculated and discussed.

\section{Formalism}
\label{Form}

The disordered fcc and bcc phases of Al$_{x}$CrFeCoNi alloy
with $x$ ranging from $x$=0 to $x$=2 were studied for
experimentally observed phases and lattice constants \cite{axcfcn}.
The fcc phase exists roughly for $x < 0.5$ while the
bcc phase is stable for $x > 1.0$, but boundaries are not
well defined.
Duplex phase (a mixture of fcc and bcc phases) exists for the
intermediate Al concentrations.
We note that sometimes the Al$_{x}$CrFeCoNi alloy is presented
as Al$_{1-4y}$Cr$_{y}$Fe$_{y}$Co$_{y}$Ni$_{y}$, where $y=1/(4+x)$
and the sum of all component concentrations is one \cite{hea-rev}.

The spin-polarized electronic structure calculations were
done using the Green function formulation of the tight-binding
linear muffin-tin orbital (TB-LMTO) method in the atomic sphere
approximation (ASA) \cite{book}.
We employ the scalar-relativistic version of the TB-LMTO method
and, in order to assess the importance of the relativistic effects,
we also made calculations using the fully relativistic version
of the TB-LMTO method.
In both cases the exchange-correlation potential of Vosko, Wilk
and Nusair (VWN)\cite{VWN} and the $spd$-basis set were used.
The alloy disorder in studied multicomponent alloys is
described in the framework of the coherent potential approximation
(CPA) \cite{lmtocpa}.
The use of the CPA allows us to work very efficiently in small
fcc or bcc unit cells.
On the contrary, large special-quasirandom structure (SQS)
supercells are needed in conventional density-functional-theory (DFT)
studies \cite{cantor,cantor2}.
It should be noted that continuously varying Al content imposes
additional non-trivial constraints on the choice of a suitable
SQS-supercell.
The cost we pay for using the CPA is the neglect of possible
local environment and clustering effects in the alloy which,
on the other hand, are captured by the SQS-supercell approach.
The CPA is a reliable approach in well disordered alloys,
particularly when the concentration trends are concerned.
Even more important advantage of the CPA is the fact that it
provides naturally transport relaxation times which need not
be taken from outside like, e.g., in the Boltzmann equation
approach.

The transport properties are described by the conductivity tensor
$\sigma$ with components $\sigma_{\mu \nu}$ ($\mu,\nu$=x,y,z).
The resistivity tensor $\rho$ with components $\rho_{\mu \nu}$
is obtained simply by inversion of the conductivity tensor,
$\rho=\sigma^{-1}$.
The conductivity tensor is determined in the framework of the
Kubo-Greenwood (K-G) approach (only diagonal elements of
$\sigma_{\mu \nu}$ are non-zero in present cubic systems in the
scalar-relativistic model \cite{sigma}).
The off-diagonal components of $\sigma_{\mu \nu}$ are needed for
the AMR/AHR studies and they are calculated in the framework of the
Kubo-Bastin (K-B) formulation of the fully-relativistic transport
in disordered magnetic alloys which includes both the Fermi-surface
and Fermi-sea terms on equal footing \cite{rsigma}.
The Fermi-surface term contains contribution only from the
states at the Fermi energy and includes the most important
elastic scattering effects due to impurities.
The Fermi-sea term, on the contrary, depends on all occupied states
below the Fermi energy; this term contributes only to the
antisymmetric part of the tensor $\sigma_{\mu \nu}$.
Once the transport tensor is determined, the AHR=$\rho_{\rm xy}$ while
the AMR=($\rho_{\rm zz}-\rho_{\rm xx})/\rho_{\rm tot}$, where $\rho_{\rm tot}$
is the average value of diagonal components of the resistivity
tensor.
In relativistic calculations we assume that the magnetic moment
points in the z-direction.
The disorder-induced vertex corrections \cite{vertex}, which describe
the correlated motion of two electrons in a random alloy potential,
are included.
They correspond to the backward scattering contribution in the
conventional Boltzmann equation approach.

The Gilbert damping (GD) constant is an important phenomenological
parameter describing the magnetization dynamics.
It is evaluated here with the help of a recently developed
approach using nonlocal torques\cite{gd-our} as an alternative
to the usual local torque operators entering the
torque-correlation formula.\cite{gd-starikov,gd-sakuma,gd-kkr}
This leads to effective torques that are represented as
non-site-diagonal and spin-independent matrices, which
simplifies evaluation of disorder-induced vertex corrections
which play essential role in the present formulation since their
neglect would lead to quantitatively and physically incorrect
results.\cite{gd-our}

Our formulation gives results that compare well to other
first-principles studies.\cite{gd-starikov,gd-sakuma,gd-kkr}
In this study we will concentrate on the GD due to chemical
disorder, especially the effect of Al-doping.
It should be noted that there are other sources of damping, e.g.,
the temperature effects due to phonons and spin fluctuations
which are neglected here.

\section{Results}
\label{Res}
\subsection{Electronic structure and magnetic moments}
\label{ES}

The results of electronic structure calculations serve as an input
for transport calculations of Al$_{x}$CrFeCoNi alloys.

\begin{figure}[h]
\center \includegraphics[width=7.5cm]{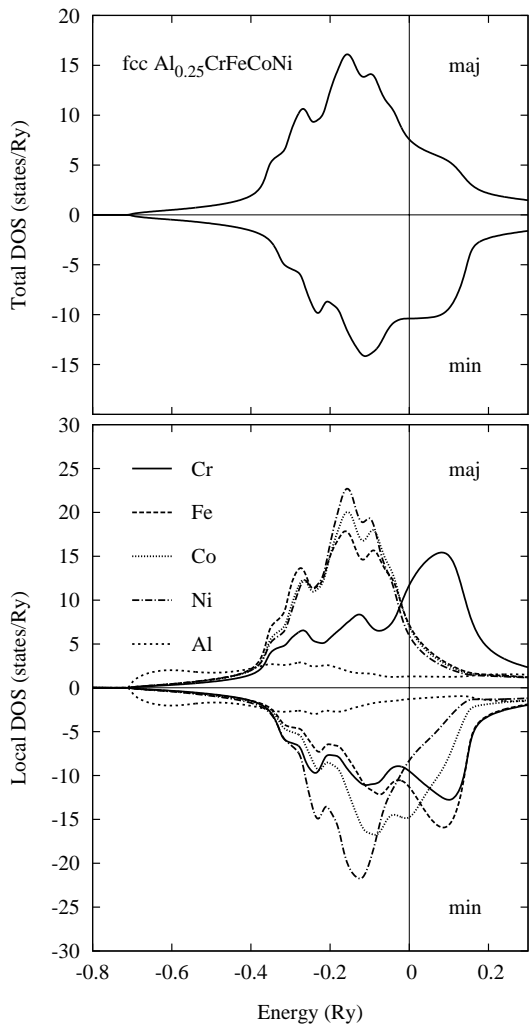}
\caption { Calculated spin-resolved DOS's for fcc Al$_{0.25}$CrFeCoNi
alloy: the total DOS (upper frame) and atom-resolved DOS's (lower
frame) are shown. The vertical lines denote the position of the
Fermi level.
}
\label{f1}
\end{figure}

\begin{figure}[t]
\center \includegraphics[width=7.5cm]{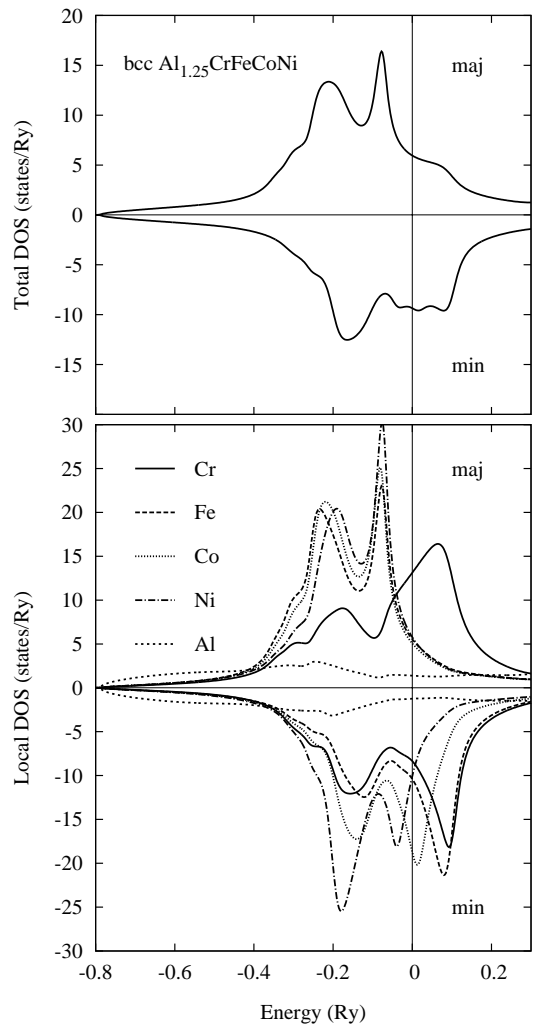}
\caption { The same as in Fig.~1 but for
bcc Al$_{1.25}$CrFeCoNi alloy.
}
\label{f2}
\end{figure}

To illustrate the underlying electronic structure, we show in
Figs.~1 and 2 the total and component-resolved densities of
states (DOS) for two typical alloys, namely, fcc Al$_{0.25}$CrFeCoNi
and bcc Al$_{1.25}$CrFeCoNi.
The following conclusions can be done:
(i) We note a typical two-peak-like total DOS characteristic of
the bcc phase as compared to an essentially one-peak-like total DOS
for the fcc phase.
In both cases the Fermi level is located deep inside the valence band
as it is typical for metallic systems in contrast, e.g.,
to doped semiconductors in which the clustering has a non-negligible effect
on DOS close to band edges.
The CPA is thus a good approximation for electron transport studies;
(ii) We also note larger total DOS at the Fermi energy for fcc phase
as compared to the bcc phase. This indicates a larger amount of
carriers and thus a smaller resistivity/larger conductivity of
fcc phases because the amount of disorder is similar in both
phases (see atom-resolved DOS's below and discussion there);
(iii) The Al-resolved DOS is free-electron like with only small
modifications in the energy region where it hybridizes with
transition metal states;
(iv) The majority Ni-, Co-, and Fe-resolved DOS's indicate only
negligible influence of the disorder: they all have similar
shapes and centers of gravity and resemble corresponding total
DOS.
On the contrary, the majority Cr-DOS has the center of
gravity shifted to higher energies and its shape is different
from those of Ni-, Co-, and Fe-states.
This is due to a lower atomic number of Cr and thus a weaker Coulomb
attraction in comparison with Fe, Co, and Ni.
Majority Cr states thus introduce a significant disorder in the
majority band.
Below we show that in present alloys the resistivity in both
majority and minority channels is comparable (see also
Ref.~\onlinecite{nifeco});
(v) The minority Ni-, Co-, Fe-, and Cr-resolved DOS's have
centers of gravity shifted to different positions thus indicating the
presence of disorder among all components as contrasted to
the majority states.
It should be noted that the character of disorder in both
fcc and bcc phases is quite similar.
The presence of non-negligible disorder in both majority and
minority states is the reason of much larger resistivity of
Al$_{x}$CrFeCoNi alloys as compared to the Ni-rich NiFe and
NiCo alloys in which the disorder effect in majority spin
channel is negligible \cite{nifeco}.
The disorder effect is essential for both, minority and majority
spin channels, but operates in them differently.
This observation provides us with a motivation to study the
magnetotransport phenomena.

\begin{table}[h]
\caption{Calculated total magnetic moment ($M_{\rm tot}$) and local
magnetic moments ($m^{\rm X}$, X=Al, Cr, Fe, Co, Ni) for
Al$_{x}$CrFeCoNi alloys in the fcc ($x_{\rm Al}=0.25$) and
bcc ($x_{\rm Al}=1.25$) phases. Magnetic moments are in $\mu_{\rm B}$.
}
\begin{tabular}{|c||c|c|c|c|c|c|} \hline
 $x_{\rm Al}$ & $M_{\rm tot}$ & $m^{\rm Al}$ & $m^{\rm Cr}$ & $m^{\rm Fe}$ &     $m^{\rm Co}$ & $m^{\rm Ni}$ \\
\hline
 0.25 (fcc) & 0.606 & $-$0.054 & $-$0.620 & 1.933 & 1.027 & 0.253 \\
 1.25 (bcc) & 0.646 & $-$0.045 & $-$0.103 & 2.117 & 1.243 & 0.191 \\
\hline
\end{tabular}
\label{t1}
\end{table}

We present magnetic properties in Table~1, where we show the
total and local magnetic moments for alloys with $x=0.25$ and
$x=1.25$.
We can make the following conclusions:
(i) The induced local Al moments are very small and negative. Also
moments on Cr atoms are negative and their absolute value is
reduced with increasing Al content. The present alloys are
thus ferrimagnets.
The values of Ni-local moments are strongly reduced as compared
to the fcc Ni crystal;
(ii) Dominating moments are those on Co and, first of all, on
Fe sites which have values close to the values in bcc Fe crystal while
moments on Co-sites are smaller as compared to fcc-Co crystal.
Both moments depend weakly on the Al doping; and
(iii) Due to the character of local moments, both alloys
have non-zero total magnetization with total moments slightly larger
for the bcc phase and relatively small in their sizes, being of
order 0.5~$\mu_{\rm B}$.
We note a good quantitative agreement of present moments with a
recent theoretical study \cite{js-axcfcn}.

The present CPA calculations ignored any spin
fluctuations in the ground state of the alloys. However,
existing studies of the Cantor CrMnFeCoNi alloy \cite{hea-ebert}
and of the ternary fcc NiCoMn system \cite{Mu-PRMater} revealed
an instability of Mn atoms to form more complicated moment distributions;
we have verified this feature for the quinary Cantor  alloy by using
the well-known CPA approach \cite{Akai1993}.
In order to examine a similar instability of Cr atoms in the present
Al$_x$CrFeCoNi systems, we have performed the CPA calculations which
started with multiple Cr magnetic moments.
For both structures (fcc and bcc), the iterations converged always to
the same single value of Cr moment. This indicates that the present
AlCrFeCoNi systems can be reasonably described by assuming the same
(average) local moment attached to each alloy species.

\subsection{Residual resistivities}
\label{Rho}
The theoretical estimate of residual resistivity $\rho_0$ and its
comparison with available experiments \cite{axcfcn,rho-exp,axcfcn2}
is the main result of the present paper.
We treat the fcc and bcc phases as disordered alloys described by the CPA.
Corresponding electronic structure provides also naturally the
transport relaxation times as used in the K-G formula for estimate
of resistivities.
We note that the SQS-supercell calculations (see, e.g.,
Refs.~\onlinecite{cantor,cantor2}) indicate the presence of local environment
effects, both in atomic structure and spins, which are neglected here.
On the other hand, local environment effects appear as fluctuations
around average atomic levels.
Although such fluctuations are large in some cases, they usually
correspond to states with small weights.
One can thus say that the intrinsic chemical disorder due to many
atomic components will dominate.
Considering further the fact that the CPA gives reliably
concentration trends, we can conclude that present approach
represents reasonable first approximation to estimate resistivities
in the present HEA's.

\begin{figure}
\center \includegraphics[width=7.0cm]{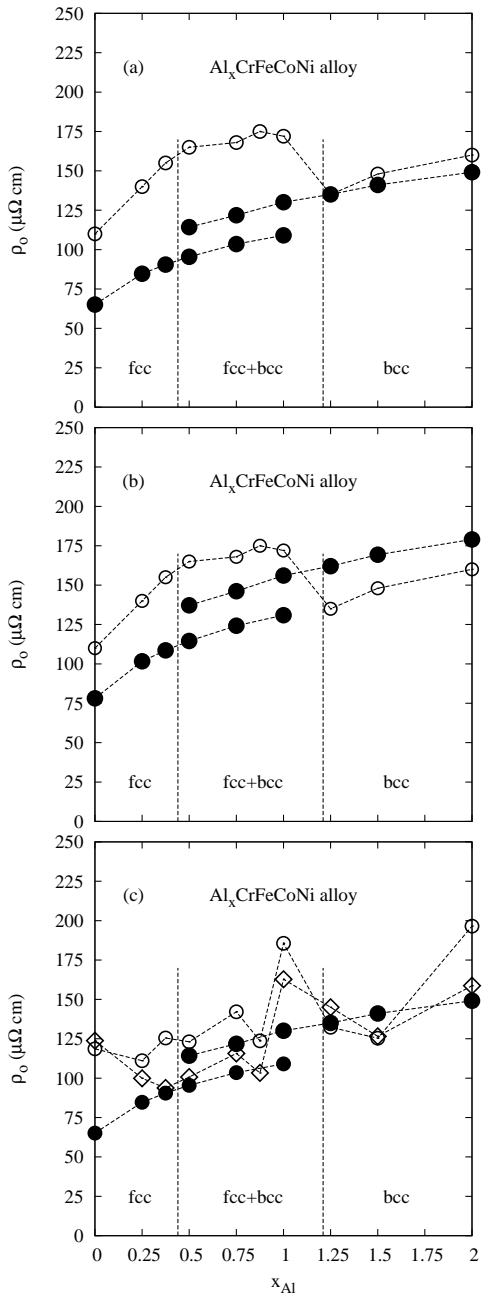}
\caption {Calculated and experimental residual resistivities
$\rho_0$ as a function of Al concentration $x_{\rm Al}$.
Dashed vertical lines denote approximate boundaries between the
fcc, duplex, and bcc phases.
Note that the fcc phase extends into duplex phase (fcc + bcc)
and similarly the bcc phase starts in duplex phase.
Filled circles denote theoretical results while empty symbols
correspond to experimental data.
(a) Experiment Ref.~\onlinecite{axcfcn} at room temperature compared
with theoretical results for $T = 0$ K.
(b) The same as (a) but theoretical results are scaled by a
factor 1.2 to fit experiment done at room temperature (see text).
(c) Comparison of experiment, Ref.~\onlinecite{axcfcn2}, with
theoretical results.
Empty circles and empty boxes denote alloys 'as-cast' and 'homogenized',
respectively (see experimental paper for details).
Both the theory and experiment now correspond to $T=0$~K.
}
\label{f3}
\end{figure}

Results are shown in Fig.~3 for both fcc and bcc Al$_{x}$CrFeCoNi
alloys for which experiments \cite{axcfcn,axcfcn2} are available.

Let us start with experiment Ref.~\onlinecite{axcfcn} (see also
Ref.~\onlinecite{rho-exp}).
It should be noted that experiment was done at the room temperature
while calculations relate to $T=0$~K.
We note an enhancement of the resistivity due to the lattice vibrations
and spin fluctuations induced by a finite temperature.
It is possible to include, for simple systems, the effect of temperature
in the framework of the alloy analogy model as formulated in the CPA
\cite{rhoT}.
It should be noted that although a success was recently reached
by a modified approach for fcc alloys with few alloy constituents
\cite{rhoT2} such a detailed study of temperature effects is
beyond the scope of the present paper.
Under such situation we decided just to scale the zero temperature
results by an empirical constant to account for the finite
temperature effects.
We have chosen this factor to be 1.2 motivated by the experiment
\cite{axcfcn2} in which resistivity ratio for 300~K to 0~K was
measured for many samples of different compositions.
The purpose is just to see the effect of the finite temperature.
Results are shown in Figs.~3a,3b.
Already large values of $\rho_0$ exist for $x_{\rm Al} = 0.0$,
i.e., for the equiconcentration quaternary CrFeCoNi alloy, which
agrees very well with a recent theoretical study \cite{hea-ebert}.
We observe an increase of $\rho_0$ with increasing Al content
in both the fcc and bcc phases. This result as well as large values
of $\rho_0$ are due to the fact that $d$-states of Al are missing
and thus Al has a low density of states around the Fermi energy
in comparison with other alloy components introducing thus a strong
scattering.
For example, in disordered bcc Fe$_{1-x}$Al$_x$
alloys \cite{feal} the experimental $\rho_0$ is about
150~$\mu\Omega$cm at $T$=4~K for $x_{\rm Al}$ around 0.3.
Similarly, the random bcc V$_{0.75}$Al$_{0.25}$ alloy exhibits
practically the same resistivity \cite{val-all}.
Comparable values are obtained for Al$_{x}$CrFeCoNi for large
$x_{\rm Al}$ in the bcc phase.

There is a good quantitative agreement of calculated and
measured $\rho_0$ in both fcc and bcc regions, in particular
for the scaled model. In agreement with the experiment the
slope of the concentration dependence of $\rho_0$ is larger for
the fcc phase as compared to the bcc one, although the effect
is more pronounced in the experiment.

It is a well-known fact \cite{nifeco} that a significant
increase of the resistivity occurs in Ni-rich NiFe and NiCo fcc
alloys due to the mixing of spin-channels by the spin-orbit coupling.
We have therefore performed also fully-relativistic calculations
for $x_{\rm Al} = 0.25$ and $x_{\rm Al} = 1.25$ alloys with fcc and
bcc phases, respectively.
We have obtained only a small changes of $\rho_0$, namely,
$\rho_0$ was 86.16~$\mu\Omega$cm vs 84.70~$\mu\Omega$cm for
$x_{\rm Al} = 0.25$, and 137.42~$\mu\Omega$cm vs 135.03~$\mu\Omega$cm
for $x_{\rm Al} = 1.25$.
In both cases, higher values correspond to the fully-relativistic model.
The origin of large enhancement of $\rho_0$ by spin-orbit coupling in the
above-mentioned binary alloys is the existence of disorder-free majority
bands, which are missing here.
The contributions of the spin-up ($\sigma^{\uparrow}$) and spin-down
($\sigma^{\downarrow}$) conductivity channels to the total conductivity
are comparable.
For example,
$\sigma^{\uparrow}$ ($\sigma^{\downarrow}$) = 4.77 (7.04)~kS/cm
for $x_{\rm Al} = 0.25$, and
$\sigma^{\uparrow}$ ($\sigma^{\downarrow}$) = 3.85 (3.56)~kS/cm
for $x_{\rm Al} = 1.25$, respectively.
We can thus conclude that relativistic effects are small in the studied
alloy.

Theoretical description of the duplex region in which both phases
co-exist is difficult because of the lack of structural details.
We therefore present separately results  for the fcc phase
with Al concentrations extending into the duplex (fcc+bcc) region,
and similarly, we start the bcc phase in the duplex region.
Corresponding lattice constants were taken from  Ref.~\onlinecite{axcfcn}.
We have thus avoided any processing of results like, e.g., the serial
or parallel resistivities, the arithmetic weighting, etc.
Inhomogeneity of samples in the duplex region is obvious.

Recently Singh et al. \cite{Singh} have shown on the basis of
calculations of chemical interatomic interactions that considered
Al$_{x}$CrFeCoNi alloys can exhibit a tendency to the clustering
in the fcc phase while in the bcc phase can exist ordering tendency.
One can thus speculate that this can be one of possible reasons for
the smaller/larger calculated resistivities for fcc/bcc phases as
compared to the experimental ones .

Next we will discuss the experiment of Ref.~\onlinecite{axcfcn2}.
Results are presented in Fig.~3c with the following comments:
(i) Contrary to the experiment \cite{axcfcn} there is no clear
concentration trend. Fluctuating values of $\rho_0$ may indicate
sample inhomogeneity due to its preparation. Largest
fluctuations are, as expected, in the duplex region.
Larger fluctuations are also for 'as-cast' samples as compared to
'homogenized' ones; and
(ii) Nevertheless, calculated and measured resistivity values
are still in acceptable agreement, as well as larger resistivities
for higher $x_{\rm Al}$ (bcc phase).
Experiment gives no detailed structural data concerning
studied samples, just its Al-content so that more detailed
discussion of measured resistivity fluctuations and their
comparison with the experiment is not possible.

We note that the performed calculations of residual resistivity ignore
the effect of local atomic relaxations, which provide an additional
mechanism of electron scattering.
Since we cannot determine the magnitude of the local relaxations
quantitatively, we have performed only a preliminary study in order to
get rough estimation of their effect on the resistivity by employing
the alloy analogy model in the CPA \cite{rhoT, Wagenknecht2019}.
As a typical mean value of the atomic displacement, we took $\Delta u
= 0.05 $ \AA ~for fcc alloys (obtained for the fcc Cantor alloy
\cite{hea-ebert}) and a slightly higher value $\Delta u = 0.075 $ \AA
~for bcc alloys (because of the more open bcc geometry).
The resulting increase of the residual resistivity of Al$_x$CrFeCoNi
was small in both structures, being about 2.8~\% for $x=0.25$ in the
fcc case and about 1.4~\% for $x=1.25$ in the bcc case.
These results agree qualitatively with those of
Ref.~\onlinecite{hea-ebert} proving the dominating effect of strong
intrinsic chemical disorder on the resistivity.
A more systematic study of the role of local atomic relaxations
goes beyond the scope of the present work.

One can summarize that the CPA, despite of its simplicity,
is able to reproduce the main features of measured resistivities
also in such complex alloys like Al$_{x}$CrFeCoNi.
Clearly, the main reason for this success is the dominance of
intrinsic chemical disorder in alloy. On the other hand, one should
keep in mind that calculated values are influenced by the neglect of
lattice relaxations.
In general, one could say that lattice relaxations roughly represent
site-off diagonal disorder which have much smaller effect as compared
to the dominating chemical disorder related to different positions of
atomic alloy levels.

\subsection{Spin-disorder resistivity (SDR)}
\label{SDR}

The SDR is the resistivity caused by spin fluctuations
that exist at finite temperature in the paramagnetic state above
the Curie temperature. The local moments still exist but they are
oriented randomly in such a way that the total magnetic moment
is zero.
 From the theoretical point of view the SDR can be simulated
successfully in the framework of the CPA as the resistivity of
an equiconcentration alloy of spin moments pointing in opposite
directions (the disordered local moment (DLM) state \cite{dlm}).
The fluctuating local moments are then determined selfconsistently
in the framework of the DFT.
In the fcc region only local moments on Fe-sites are nonzero, all
other collapse to zero.
In the bcc region, in addition, local moments on Co atoms survive.
Such result, in general, is not correct as, e.g., the local DLM
moment in fcc Ni collapses to zero but the experiment indicates
its nonzero value at the Curie temperature.
These values can be found theoretically not only for fcc Ni \cite{sdr},
but also for binary alloys \cite{sdr2}.
The situation is much more complicated for the present
multicomponent alloy.
We therefore determine just the lower and upper limits of the SDR.
The lower limit is the above DLM result, the upper limit corresponds
to the DLM state which is constructed on the basis of an FM solution
assuming the frozen Fermi energy and frozen potential parameters
\cite{sdr}.
It is denoted as $\rho_{\rm max}^{\rm SDR}$.

\begin{table}[h]
\caption{The spin disorder resistivity (SDR, the resistivity
due to spin fluctuations in the paramagnetic state) of
Al$_{x}$CrFeCoNi for two values of Al concentrations, namely,
$x_{\rm Al}=0.25$ (fcc) and $x_{\rm Al}=1.25$ (bcc) are shown.
We present the SDR results for two models, one in which the
SDR is identified with the resistivity of the DLM state
($\rho_{\rm DLM}^{\rm SDR}$) and the other ($\rho_{\rm max}^{\rm SDR}$)
in which the DLM state is constructed from the corresponding FM
solutions with frozen Fermi energies and frozen potential parameters.
For a comparison we also show conventional resistivities
($\rho_{\rm FM}$, see Fig.~3) and resistivities of non-magnetic
phases ($\rho_{\rm NM}$). All values are in $\mu\Omega$cm.
}
\begin{tabular}{|c||c|c|c|c|} \hline
 $x_{\rm Al}$ & $\rho_{\rm FM}$ & $\rho_{\rm NM}$ & $\rho^{\rm SDR}_{\rm DLM}$ &
$\rho^{\rm SDR}_{\rm max}$ \\
\hline
 0.25 (fcc)  &  84.70  &  72.16 &  83.98  & 89.97 \\
 1.25 (bcc)  & 135.03  & 117.17 & 132.17  & 137.49 \\
\hline
\end{tabular}
\label{t2}
\end{table}

Results for two Al concentrations are summarized in
Table~2 in which we have added for a comparison also resistivities
of the reference FM state and resistivities of corresponding
non-magnetic phases.
We have following comments:
(i) The non-magnetic phases have slightly smaller resistivities
as compared not only to the DLM phases but also as compared to
the reference FM phase.
The effect of magnetic scatterings is thus less relevant than
the effect of different atom types and their different potentials;
(ii) Slightly larger values of the reference $\rho_{\rm FM}$ as
compared to $\rho^{\rm SDR}_{\rm DLM}$ are due to the fact that
in the DLM state in the fcc/bcc phase are non-zero only Fe and
perhaps also Co moments.
Missing magnetic scattering thus leads to smaller resistivities
(see also discussion in (i)); and
(iii) On the contrary, the $\rho^{\rm SDR}_{\rm max}$ is slightly
larger due to the presence of fluctuating moments on Cr, Fe, Co,
and Ni atoms.
One can thus conclude that due to already large resistivity of
the reference FM state, the spin disorder influences resistivity
only weakly.

\subsection{Anisotropic magnetoresistance, anomalous Hall
resistivity, and Gilbert damping}
\label{AHR}

We calculate further quantities which are due to the spin-orbit
coupling, namely, the AMR and the AHR for two typical
Al-concentrations, namely, $x_{\rm Al}$ = 0.25 (fcc phase)
and $x_{\rm Al}$ = 1.25 (bcc phase).
The relativistic input is needed to solve the K-B transport
equation \cite{rsigma}.
While we have found no experimental data for the AMR, the
AHR data are available for the above two alloys \cite{axcfcn}.

\begin{table}[h]
\caption{Calculated AMR and AHR for Al$_{x}$CrFeCoNi alloys in the
fcc ($x_{\rm Al}=0.25$) and bcc ($x_{\rm Al}=1.25$) phases.
The AMR values are in \% while the AHR values are in $\mu\Omega$cm.
}
\begin{tabular}{|c||c|c|c|} \hline
 $x_{\rm Al}$ & AMR & AHR$_{\rm th}$ & AHR$_{\rm exp}$ \\
\hline
 0.25 (fcc) & 0.031 & 0.879 & 0.5 \\
 1.25 (bcc) & 0.044 & 1.699 & 1.5 \\
\hline
\end{tabular}
\label{t3}
\end{table}

Calculated results are summarized in Table~3 with the
following conclusions:
(i) The AMR is positive, but its values are very small,
considering the fact that, e.g., for Ni-rich NiFe the AMR
can be as large as 15\% \cite{nifeco,SPSW}.
It was shown that large values of the AMR in fcc Ni-rich
alloys are due to essentially disorder-free majority bands.
On the contrary, the Ni-rich NiMn alloy has disorder in both
the majority and minority bands and significantly lower AMR than
Ni-rich NiFe, but still few times larger than the present alloys.
In addition to very similar disorder in both channels in present
alloys, the other reason of such small AMR can be the ferrimagnetic
rather than the FM character of present alloys with the
antiparallel Cr moments. Its role plays also small total moment
of studied HEA alloys; and
(ii) There is a good agreement between calculated and
measured AHR for the bcc phase ($x_{\rm Al}$ = 1.25) while the
agreement for the fcc phase ($x_{\rm Al}$ = 0.25) is worse but
still reasonable.
We note that a good agreement of both calculated resistivities
and AHR with experiment is a non-trivial result.

We have estimated GD parameters for the same typical
concentrations as above for the AHR.
Calculated values of the GD parameter for fcc ($x_{\rm Al}=0.25$)
and  bcc ($x_{\rm Al}=1.25$) phases are, respectively, 0.00655
and 0.00585.
Both values are similar which is compatible with similar
values of the DOS at the Fermi level and the total magnetization
whose ratio is a rough estimate of the GD parameter, which explains
also rather large values due to small total spin moments in
both alloys.
Calculated values of GD parameter are comparable to those
in Ni-rich fcc NiFe alloy but are larger as compared to
bcc-FeCo alloy.\cite{gd-our}

\section{Conclusions}
\label{Conc}

Transport properties of the fcc and bcc phases of
the high-entropy Al$_{x}$CrFeCoNi alloys were calculated
over a broad range of Al concentrations using the DFT-based
simulations.
The main conclusions from numerical studies can be
summarized as follows:
(i) The agreement of calculated residual resistivities with
available experimental data is good for both fcc and bcc
phase.
In particular, the resistivity values as well
as larger resistivity of the bcc-phase as compared to the
fcc one agree with both experiments.
Calculation even reproduce details of concentration trends
in one of the experiment \cite{axcfcn}.
(ii) The major contribution to the residual resistivity is
due to the intrinsic chemical disorder while the magnetic
disorder has smaller effect. The increase of $\rho_0$ with
increasing Al concentration in both fcc and bcc phases and
its large values in particular in the latter one are due to
strong scatterings on Al atoms;
(iii) The calculated values of anisotropic magnetoresistance
are positive but very small
being less than 0.05\% for both fcc and bcc phases;
(iv) The spin disorder influences resistivity only weakly because of
already large resistivity of the reference FM state;
(v) Estimated values of the Gilbert damping are comparable
for chosen typical fcc and bcc phases and are rather large
(of order 0.006) due to small total spin moments;
and
(vi) The estimated anomalous Hall resistivity again agrees well for the bcc phase
while agreement with the experiment for the fcc phase is
worse though still acceptable.

The present results thus suggest that the CPA captures the main
scattering mechanism due to intrinsic alloy disorder and gives
acceptable description even for such complex alloys like the
studied one.\\

\noindent {\bf Acknowledgments}

The work of J.K., V.D, F.M., and I.T. was supported by a Grant
from the Czech Science Foundation (No. 18-07172S) and S.K.
thanks for support from the Center for Computational Materials
Science, Vienna University of Technology.
We acknowledge the support from the National Grid Infrastructure
MetaCentrum (project CESNET LM2015042) and the Ministry of Education,
Youth and Sports (project LM2015070).

\end{document}